\documentclass{aip-cp}

\usepackage[numbers]{natbib}
\usepackage{rotating}
\usepackage[usenames,dvipsnames]{color}

\usepackage{multirow}

\begin{document}

% Title portion 
% Length <12 pages

\newcommand{\red}[1]{\textcolor{red}{#1}}
\newcommand{\bsub}{\begin{subequations}}
\newcommand{\esub}{\end{subequations}}
\newcommand{\beq}{\vspace{0.5em}\begin{equation}}
\newcommand{\eeq}{\end{equation}\vspace{0.5em}}
\newcommand{\beqn}{\vspace{0.5em}\begin{eqnarray}}
\newcommand{\eeqn}{\end{eqnarray}\par\vspace{0.5em}\noindent}
\newcommand{\br}{{\mathbf{r}}}
\newcommand{\bq}{{\mathbf{q}}}
\renewcommand{\vec}[1]{\mbox{\boldmath $#1$}} 

\title{Relativistic Mean-Field  and Beyond Approaches for Deformed Hypernuclei}

\author[aff1]{J. M. Yao} 
\author[aff2,aff3]{H. Mei}
\author[aff2,aff4]{K. Hagino}
\author[aff5,aff6]{T. Motoba}
%\eaddress{anotherauthor@thisaddress.yyy}

\affil[aff1]{FRIB/NSCL, Michigan State University, East Lansing, Michigan 48844, USA } 
\affil[aff2]{Department of Physics, Tohoku University, Sendai 980-8578, Japan}
\affil[aff3]{School of Physical Science and Technology, Southwest University, Chongqing 400715, China}
\affil[aff4]{Research Center for Electron Photon Science, Tohoku University, 1-2-1 Mikamine, Sendai 982-0826, Japan}
\affil[aff5]{Laboratory of Physics, Osaka Electro-Communication University, Neyagawa 572-8530, Japan}
\affil[aff6]{ Yukawa Institute for Theoretical Physics, Kyoto University, Kyoto 606-8502, Japan}
%\corresp[cor1]{Corresponding author: yaoj@frib.msu.edu}

\maketitle

\begin{abstract}
 We report the recent progress in relativistic mean-field (RMF) and beyond approaches for the low-energy structure of deformed hypernuclei. 
We show that the $\Lambda$ hyperon with orbital angular momentum $\ell=0$ (or $\ell>1$) generally reduces (enhances) nuclear quadrupole collectivity. 
The beyond mean-field studies of hypernuclear low-lying  states  demonstrate that  there is generally a large configuration mixing between the two components $[^{A-1}Z (I^+) \otimes \Lambda p_{1/2}]^J$ 
and $[^{A-1}Z (I\pm2 ^+) \otimes \Lambda p_{3/2}]^J$  in the hypernuclear $1/2^-_1, 3/2^-_1$ states.  The mixing weight increases as the collective correlation of nuclear core becomes stronger. Finally, we show how the energies of hypernuclear low-lying states are sensitive to parameters in the effective $N \Lambda $ interaction, the uncertainty of which has a large impact on the predicted maximal mass of neutron stars.

%Considering the fact that the uncertainty in the effective $\Lambda N$ interaction   generates a large uncertainty in the predicted maximum mass of neutron stars with the RMF approaches, we propose to put more efforts on the calibration of the effective $\Lambda N$ interaction before discussing  the so-called ``hyperon puzzle"  issue in the RMF approaches.  

\end{abstract}

% Head 1
\section{INTRODUCTION}

Hypernuclei provide a natural and accessible laboratory to study nucleon-nucleon ($NN$) and nucleon-hyperon ($NY$)  interactions in nuclear medium, the knowledge of  which is important for understanding neutron stars \cite{Gal16,Glendenning00}.   Lots of efforts have been devoted into parameterization of the bare $NN$, $NY$ interactions based on available scattering data and/or the results from lattice QCD calculations.  These bare interactions have been implemented into few-body calculations for very light hypernuclear systems. For heavier (hyper)nuclear systems,  the situation becomes much more complicated in the sense that a much  larger model space and thus more expensive calculations are demanded to achieve convergence in the solutions. To overcome this difficulty, one may employ the techniques of G matrix \cite{Hao93,Schulze95}, many-body perturbation theory or (in-medium) similarity renormalization group \cite{Wirth18} to dilute the resolution of the interactions and/or to build many-body correlations into the interactions. With these treatments, the realistic $NN, NY$ interactions are transformed into effective interactions that are more suitable for the studies with less expensive nuclear models. 

Alternatively, instead of starting from the bare interactions, the self-consistent mean-field or energy density functional (EDF) approaches start from an effective interaction or a universal EDF with their parameters determined directly from the properties of nuclear many-body systems. The effective interactions which already include many-body correlations from the beginning can reproduce excellently the saturation properties of nuclear matter at the mean-field level and therefore turn out to be very successful in the mean-field studies of both ordinary nuclei and hypernuclei ranging from light to heavy mass regions for several decades \cite{Bender03}. Within this framework, some interesting phenomena related to hyperon impurity effect on atomic nuclei have been disclosed, such as the shrinkage of nuclear size and the extension of nucleon driplines \cite{Vretenar99}.   

The relativistic mean-field  (RMF) or covariant EDFs approaches are of particular interest in nuclear physics as Lorentz invariance is one of the underlying symmetries of QCD \cite{Meng16}. This symmetry not only allows to describe the spin-orbit coupling of nucleons, which has an essential influence on the underlying shell structure, in a consistent way, but also puts stringent restrictions on the number of parameters in the corresponding functionals.  This character is particular important for understanding the weak hyperon spin-orbit interaction. With the RMF approaches, there have been lots of studies for spherical hypernuclei, see Ref.~\cite{Hagino16}. In recent decade, these studies  are extended into deformed hypernuclei \cite{Win08,Lu11,Xu15,Xue15} and superdeformed hypernuclei \cite{Lu14,Wu17}. The hyperon turns out to change significantly the equilibrium shape of some carbon and silicon isotopes from oblate  to spherical shape.  Considering these (hyper)nuclei are rather soft against shape fluctuations, the mean-field approximation is prone to overestimate the hyperon impurity effect on nuclear shapes \cite{Mei18}. Moreover,  the adopted effective $NY$ interactions are often optimized to fit the hyperon separation energies and the hyperon spin-orbit splitting at the mean-field level. Previous studies already showed some hints that the effective $NY$ interactions cannot be uniquely determined with these data. On the other hand,  it is also not clear if the interactions obtained in this way are applicable for hypernuclear excited states.  To clarify these questions and to make use of the  rich spectroscopic data from hypernuclear $\gamma$-ray experiments  \cite{Hashimoto06} additionally to constrain the $NY$ interaction, two beyond mean-field models for hypernuclei, namely the relativistic generator coordinate method (GCM)  \cite{Mei16R} and the relativistic particle-core coupling model  \cite{Mei14,Mei15,Mei16,Mei17}, have been established. These two models are applied to analyze the collective correlations in hypernuclear low-lying states and they extend significantly the scope of understanding hypernuclear structure from mean-field pictures.  In this contribution, we review our recent progress in the studies of  the low-energy structural properties of deformed $\Lambda$ hypernuclei within the RMF and beyond approaches. 

\section{The RMF approaches for $\Lambda$ hypernuclei}
 The Lagrangian density ${\cal L}$ for  a hypernucleus can be generally written as
  \begin{equation}
  \label{Lag1}
  {\cal L} = {\cal L}^{\rm free} + {\cal L}^{\rm em}  + {\cal L}^{\phi} + {\cal L}^{NN} + {\cal L}^{NY},
   \end{equation}
  where the Lagrangian density  for free baryons ${\cal L}^{\rm free}$ and that for electromagnetic field ${\cal L}^{\rm em}$ are
  \begin{eqnarray}
    {\cal L}^{\rm free} &=& \sum_{B=N, Y}\bar\psi^B ( i\gamma^\mu\partial_\mu - m_B)\psi^B, \\
    {\cal L}^{\rm em} &=& -\frac{1}{4}F^{\mu\nu} F_{\mu\nu} - e\bar\psi\gamma_\mu\frac{1-\tau_3}{2}\psi  A^\mu.
 \end{eqnarray}
 The $\psi^B$ represents either the nucleon ($B=N$) or hyperon ($B=Y$) field,  $m_B$ for the corresponding mass and $F^{\mu\nu}$ for the field tensors of the electromagnetic field $A^\mu$, defined as $F^{\mu\nu}=\partial^\mu A^\nu- \partial^\nu A^\mu$. The ${\cal L}^{\phi}$ term is for meson fields. The last two terms ${\cal L}^{NN}$ and ${\cal L}^{NY}$ in Eq. (\ref{Lag1}) describe the effective $NN$ and $NY$  interactions and they are parameterized phenomenologically into different forms in different version of RMF approaches. Generally speaking,  these terms can be classified into two types, i.e., the meson-exchange version with  the presence of ${\cal L}^{\phi}$ and the point-coupling version without the ${\cal L}^{\phi}$ term, according to the way how the nucleons and hyperons interact with each other in the hypernuclei. The expression for the $NN$ interaction can be found for example in Ref.~\cite{Meng16}. For the sake of simplicity, only the $\Lambda$ hyperon is considered here. The effective $N\Lambda$ interaction can be chosen as follows.
 
\begin{itemize}
\item The   Lagrangian density for the $N\Lambda $ interactions in terms of exchange effective scalar ($\sigma$) and vector ($\omega$) mesons read

\begin{eqnarray}
\label{FR-NY}
 \mathcal{L}^{N\Lambda}
 & = & \bar{\psi}^\Lambda \left(- g_{\sigma \Lambda} \sigma
                          - g_{\omega \Lambda} \gamma^{\mu} \omega_{\mu}
                          + \frac{f_{\omega \Lambda\Lambda}}{4m_\Lambda}  \sigma^{\mu\nu} \Omega_{\mu\nu}  \right)
             \psi^\Lambda,
 \label{eq:LL}
\end{eqnarray}
with the field tensor defined as $\Omega^{\mu\nu}=\partial^\mu \omega^\nu- \partial^\nu \omega^\mu$. The
 $g_{\sigma \Lambda}$ and $g_{\omega \Lambda}$ are the coupling constants of the hyperon with the scalar $\sigma$ and vector $\omega$ meson fields, respectively. 
The term proportional to $f_{\omega \Lambda\Lambda}$ with $\sigma^{\mu\nu}=\frac{i}{2}[\gamma^\mu, \gamma^\nu]$
represents the tensor coupling between the hyperon and the $\omega$ meson field. The above $\Lambda N$ interaction introduces three additional free parameters $g_{\sigma \Lambda},  g_{\omega \Lambda}, f_{\omega \Lambda\Lambda}$.

\item  The Lagrangian density for the  contact version of the $N\Lambda $ interaction can be constructed by eliminating the meson fields and expanding the meson propagators up to the next-to-leading order terms,
\begin{eqnarray} 
\label{YN-PC}
{\cal L}^{N\! \Lambda}=
{\cal L}_{\rm 4f}^{N\! \Lambda}+{\cal L}_{\rm der}^{N\!\Lambda}
+{\cal L}_{\rm ten}^{N\!\Lambda}, \end{eqnarray}
with
\begin{eqnarray}
{\cal L}_{\rm 4f}^{N\!\Lambda}
&=&
-\alpha_S^{(N\Lambda)}(\bar{\psi}^{N}\psi^N)(\bar{\psi}^{\Lambda}\psi^{\Lambda})
%\nonumber\\
%&&
-\alpha_V^{(N\!\Lambda)}(\bar{\psi}^{N}\gamma_{\mu}\psi^N)
(\bar{\psi}^{\Lambda}\gamma^{\mu}\psi^{\Lambda}), \\
{\cal L}_{\rm der}^{N\!\Lambda}
&=&-\delta_S^{(N\!\Lambda)}(\partial_{\mu}\bar{\psi}^{N}\psi^N)
(\partial^{\mu}\bar{\psi}^{\Lambda}\psi^{\Lambda}) 
-\delta_V^{(N\!\Lambda)}(\partial_{\mu}\bar{\psi}^{N}\gamma_{\nu}\psi^N)
(\partial^{\mu}\bar{\psi}^{\Lambda}\gamma^{\nu}\psi^{\Lambda}),\\
{\cal L}_{\rm ten}^{N\!\Lambda}
&=&
\alpha^{(N\!\Lambda)}_T(\bar{\psi}^{\Lambda}\sigma^{\mu\nu}\psi^{\Lambda})
(\partial_{\nu}\bar{\psi}^{N}\gamma_{\mu}\psi^N).
\end{eqnarray} 
Here are five free parameters $\alpha_S^{(N\Lambda)}, \alpha_V^{(N\Lambda)}, \delta_S^{(N\!\Lambda)}, \delta_V^{(N\!\Lambda)}, \alpha^{(N\!\Lambda)}_T$, which are related to the $g_{\sigma \Lambda},  g_{\omega \Lambda}, f_{\omega \Lambda\Lambda}$ approximately  by
 \beq
 \alpha_S^{(N\Lambda)} \approx -\frac{g_{\sigma \Lambda}g_{\sigma N}}{m_{\sigma \Lambda}^2},\quad 
  \alpha_V^{(N\Lambda)} \approx \frac{g_{\omega \Lambda} g_{\omega N}}{m_{\omega \Lambda}^2},\quad
  \alpha^{(N\!\Lambda)}_T\approx -\frac{f_{\omega \Lambda\Lambda} g_{\omega N}  }{2 m_\Lambda m^2_\omega}.
 \eeq

\end{itemize}
In literatures, the ratios $x_\sigma(=g_{\sigma \Lambda}/g_{\sigma N})$ and $x_\omega(=g_{\omega \Lambda}/g_{\omega N})$ are often introduced to define the $ N\Lambda$ interaction, relative to the $NN$ interaction. The data of $\Lambda$ binding energies in a set of hypernuclei, together with  the spin-orbit splitting of the $p$-orbital $\Lambda$ in $^{13}_\Lambda$C, are usually adopted to determine the free parameters in the effective $\Lambda N$ interactions~\cite{Keil00,Tanimura12,Wang13}.  

\section{ The beyond RMF approaches for $\Lambda$ hypernuclei}
Here we introduce two beyond-mean-field approaches for hypernuclear low-lying states, namely,  the generate coordinate method (GCM) and the particle-core coupling (PCC) or 
also called microscopic particle-rotor  model.  These two approaches are built based on the  solutions of the RMF approaches using the same $NN$ and $N\Lambda $ interactions and thus provide complementary analysis of hypernuclear low-lying states.

\begin{itemize}
\item In the GCM, the hypernuclear Hamiltonian is diagonalized in the basis formed by quantum-number projected mean-field states. 
The hypernuclear wave function is constructed as ~\cite{Mei16R}
\beq
\label{GCM:wf}
| \Psi^{JM}_{\alpha} \rangle
= \sum _{n, \beta, K} f^{J}_{n\alpha}(\beta)\hat{P}^J_{MK}\hat{P}^N \hat{P}^Z|\Phi^{(N\Lambda)}_{n}(\beta)\rangle,
\eeq
where the index $n$ refers to a different hyperon orbital state, and the index $\alpha$ labels the quantum numbers of the state
other than the angular momentum. The mean-field states $|\Phi^{(N\Lambda)}_{n}(\beta)\rangle$ are generated with
deformation constrained RMF calculations for the whole $\Lambda$ hypernuclei~\cite{Win08,Xue15}.  For simplicity, axial symmetry
is imposed and in this case, there is no $K$ mixing in the GCM state (\ref{GCM:wf}). Since the hyperon and the nucleons are not mixed,
the mean-field states can be decomposed as
\beq
|\Phi^{(N\Lambda)}_{n}(\beta)\rangle
= |\Phi^N(\beta)\rangle \otimes |\varphi^{\Lambda}_{n}(\beta)\rangle,
\eeq
where $|\Phi^N(\beta)\rangle $ and
$|\varphi^{\Lambda}_{n}(\beta)\rangle$ are the mean-field wave
functions for the nuclear core and the hyperon, respectively.

The weight function $f^{J }_{n\alpha}(\beta)$ in (\ref{GCM:wf}) 
is  determined by the variational principle, which leads to the 
Hill-Wheeler-Griffin (HWG) equation, 
 \beq
 \label{HWE}
 \sum_{\beta'}
 \left[{\cal H}^J_{n}(\beta,\beta') -E^{J}_{n\alpha}
{\cal N}^J_{n }(\beta,\beta')\right]
  f^{J}_{n\alpha}(\beta')=0,
 \eeq
 where the norm kernel 
${\cal N}^J_{n }(\beta,\beta')$ 
and the Hamiltonian kernel 
${\cal H}^J_{n }(\beta,\beta')$ 
are defined as
\beq
{\cal O}^J_{n}(\beta,\beta')\equiv
 \langle \Phi^{(N\Lambda)}_{n}(\beta) \vert \hat O
\hat{P}^J_{KK}\hat{P}^N \hat{P}^Z \vert\Phi^{(N\Lambda)}_{n}
(\beta')\rangle,
\eeq
with $\hat{O}=1$ and $\hat{O}=\hat{H}$, respectively. 
The solution of the HWG equation (\ref{HWE}) provides the energy 
$E^{J}_{n\alpha}$  and the weight function $f^{J}_{n\alpha}(\beta)$  for each of the low-lying states of hypernuclei.
It is worth mentioning that a similar beyond mean-field approach based on a Skyrme EDF for $\Lambda$ hypernuclei was developed recently \cite{Cui17}.

\item The PCC model shares the same idea of resonating-group method (RGM), in which the $\Lambda$ hypernucleus is described as a $\Lambda$ coupled to a core nucleus
~\cite{Mei14,Mei15,Mei16,Mei17}
\begin{equation}
 \label{wavefunction}
 \displaystyle \Psi^{JM}_\alpha(\vec{r},\{\vec{r}_N\})
 =\sum_{n,j,\ell, I}  {\cal R}_{j\ell I_{n}}(r)   \Big[{\cal Y}_{j\ell}(\hat{\vec{r}})\otimes
\Phi_{I_{n}}(\{\vec{r}_N\})\Big]^{(JM)}
\end{equation}
with $\vec{r}$ and $\vec{r}_N$ being the coordinate of the $\Lambda$ hyperon and that of
nucleons inside the core nucleus, respectively. $J$ is the angular momentum
for the whole system while $M$ is its
projection onto the $z$-axis.
${\cal Y}_{j\ell}(\hat{\vec{r}})$ is the spin-angular wave function for the $\Lambda$ hyperon.
$\vert\Phi_{I_{n}}\rangle$ is the wave functions of the low-lying states
of nuclear core from a GCM calculation \cite{Yao10}, where $I$ represents the angular momentum
of the core state and $n=1, 2, \ldots$ distinguish
different core states with the same angular momentum $I$. 
 For convenience, hereafter we introduce the
shorthand notation $k=\{j\ell I_{n}\}$ to represent different channels.
In contrast to the RGM for ordinary nuclear systems, there is no need to worry about the Pauli-exclusion principle between the $\Lambda$
and the nucleons inside the core.

The relative wave function ${\cal R}_{j\ell I_{n}}(r)$ of the $\Lambda$ is the radial part of a four-component Dirac spinor 
\beqn
\label{wavefunction2}
{\cal R}_{j\ell I_{n}}(r)=
\left(\begin{array}{c}
f_{j\ell I_{n}}(r) \\
i g_{j\ell I_{n}}(r)\vec{\sigma} \cdot \hat{\vec{r}}
\end{array}\right).
\eeqn
The Hamiltonian $\hat H$ for the whole $\Lambda$ hypernucleus can be written as
\beq
\hat H =\hat H_{\rm c}+ \hat T_\Lambda  + \sum^{A_c}_{i=1} \hat{V}^{(N\Lambda)}(\vec{r},\vec{r}_{N_i}),
\label{eq:H}
\eeq
where $A_c$ is the mass number of the core nucleus.  The first term in Eq. (\ref{eq:H}) is the Hamiltonian of the nuclear core, fulfilling $\hat H_{\rm c}\vert\Phi_{I_{n}}\rangle = E^I_n \vert\Phi_{I_{n}}\rangle$ and the second term $\hat T_\Lambda$ is relative kinetic energy of the $\Lambda$ hyperon. The third term represents the effect $N\Lambda $ interaction which is chosen as a contact form consistent with Eq.(\ref{YN-PC}).
Finally, one ends up with a set of coupled equations for the radial wave function
 \begin{eqnarray}
\label{couple1}
&&\left(\frac{d}{dr}-\frac{\kappa-1}{r}\right)
g_{k}(r)+(E_{I_{n}}-E_J) f_{k}(r)
+\sum_{k'}U^{kk'}_{T}(r) g_{k'}(r)
+\sum_{k'}\left[U^{kk'}_V(r)+ U^{kk'}_S(r)\right] f_{k'}(r)
=0, \\
\label{couple2}
&&\left(\frac{d}{dr}+\frac{\kappa+1}{r}\right)
f_{k}(r)-(E_{I_{n}}-2 m_{\Lambda}-E_J)g_{k}(r)
-\sum_{k'}U^{kk'}_{T}(r) f_{k'}(r)
-\sum_{k'}\left[U^{kk'}_V(r) - U^{kk'}_S(r)\right] g_{k'}(r)
= 0,
\end{eqnarray} 
where the  $\kappa$ is defined as  $\kappa=(-1)^{j+\ell+1/2}(j+1/2)$. With the multipole expansion for the $\delta(\vec{r}-\vec{r}_{Ni})$ function in coordinate space,  the vector and scalar coupling potentials in Eqs.(\ref{couple1}) and (\ref{couple2}) have the following forms
\begin{eqnarray}
\label{Coeff_couple1}
U^{kk'}_V(r)
&\equiv&\langle {\cal F}^{JM}_{k} |\alpha_V^{N\Lambda}\sum_{i=1}^{A_c}
\delta(\vec{r}-\vec{r}_{Ni})|{\cal F}^{JM}_{k'}\rangle \nonumber\\
&&=(-1)^{j'+I+J} \sum_{\lambda}  \langle j\ell  || Y_{\lambda } || j'\ell'  \rangle
\left(\begin{array}{ccc}
J       & I &  j    \\
\lambda & j'  & I'  \\
\end{array}\right) \alpha_V^{N\Lambda}\rho^{I_{n} I_{n'}}_{\lambda,V}(r),
\end{eqnarray}
\begin{eqnarray}
\label{Coeff_couple2}
U^{kk'}_S(r)
&\equiv&\langle {\cal F}^{JM}_{k} |\alpha_S^{N\Lambda}\sum_{i=1}^{A_c}\gamma^0_i
\delta(\vec{r}-\vec{r}_{Ni})|{\cal F}^{JM}_{k'}
\rangle \nonumber\\
&&=(-1)^{j'+I+J} \sum_{\lambda}  \langle j\ell  || Y_{\lambda } || j'\ell'  \rangle
\left(
\begin{array}{ccc}
J       & I &  j    \\
\lambda & j'  & I'  \\
\end{array}
\right) 
\alpha_S^{N\Lambda}\rho^{I_{n} I_{n'}}_{\lambda,S}(r),
\end{eqnarray}
and  
\begin{eqnarray}
\label{Coeff_couple3}
U^{kk'}_{T}(r)\equiv&\langle {\cal F}^{JM}_{k} |\sum_{i=1}^{A_c} \alpha_T^{N\Lambda}\left[\overleftarrow{\nabla}
\delta(\vec{r}-\vec{r}_i)+\delta(\vec{r}-\vec{r}_i)\overrightarrow{\nabla}\right] \cdot
\vec{\sigma}|{\cal F}^{JM}_{k'}\rangle.
\end{eqnarray} 

The $\rho^{I_{n} I_{n'}}_{\lambda,V}(r)$ and $\rho^{I_{n} I_{n'}}_{\lambda,S}(r)$
are vector and scalar types of reduced transition densities between nuclear core states, respectively, 
 \begin{eqnarray}
\label{TD1}
\rho^{I_{n} I_{n'}}_{\lambda, V}(r) &=&
\langle \Phi_{I_n} || 
\sum\limits_{i=1}^{A_c}
\frac{\delta(r-r_{Ni})}{r_{Ni} r}
Y_\lambda(\hat{\vec{r}}_{Ni})||\Phi_{I_{n'}} \rangle,
\\
\label{TD2}
\rho^{I_{n} I_{n'}}_{\lambda, S}(r) &=&
\langle \Phi_{I_n} || \sum\limits_{i=1}^{A_c} \gamma^0_i
\frac{\delta(r-r_{Ni})}{r_{Ni} r}
Y_\lambda(\hat{\vec{r}}_{Ni})||\Phi_{I_{n'}} \rangle.
\end{eqnarray}

\end{itemize}
  
\section{Shape polarization effect of $\Lambda$ in deformed hypernuclei}
 
\begin{figure}[h!]
\label{fig:impurity}
  \centerline{\includegraphics[width=200pt]{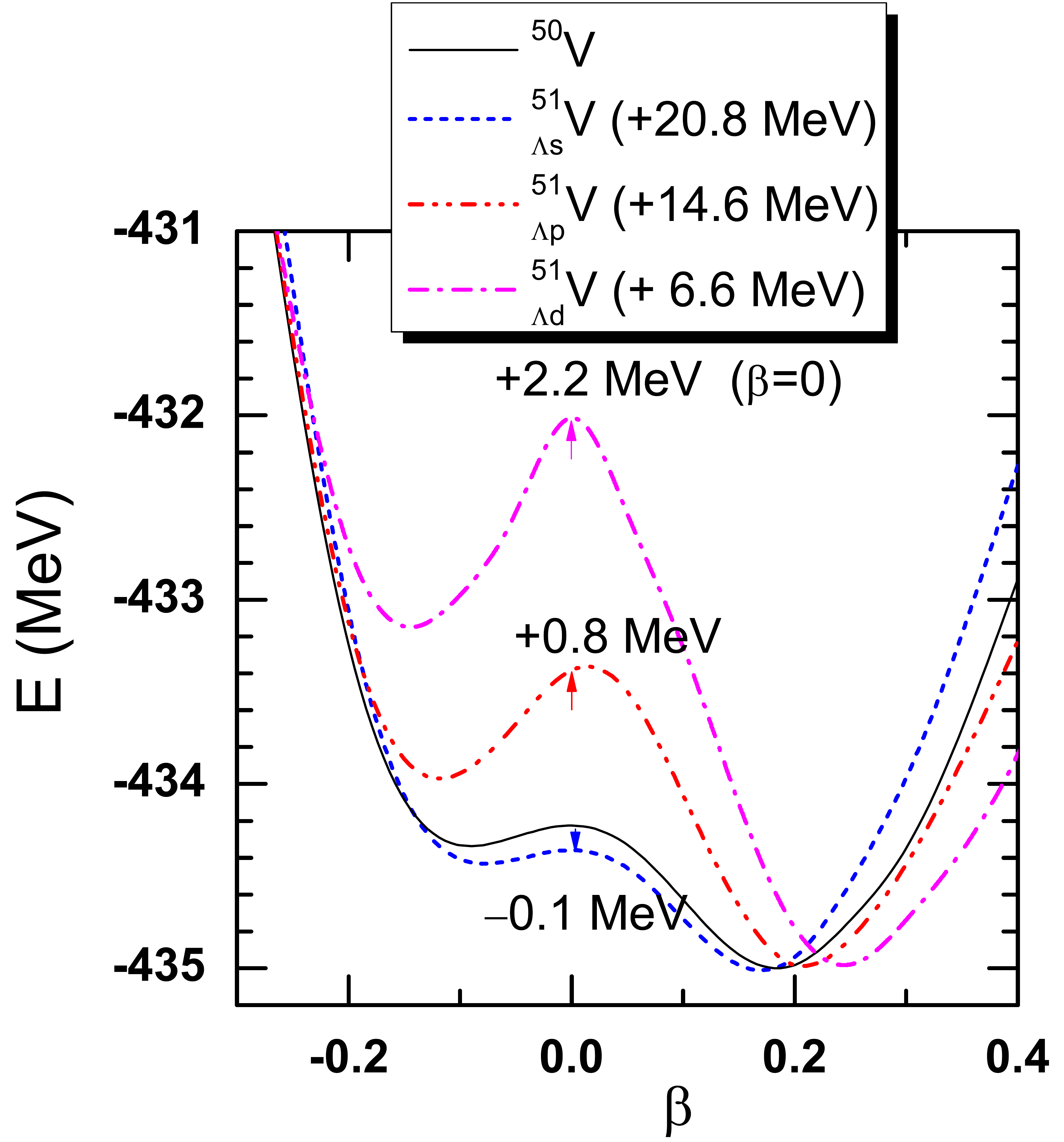}}
 \caption{(Color online) The total energy of $^{50}$V, $^{51}_{\Lambda s}$V,
  $^{51}_{\Lambda p}$V and $^{51}_{\Lambda d}$V as a function of deformation
  parameter $\beta$ from deformed RMF calculations.
  The energy of hypernuclei is shifted by normalizing the minimum
  energy to that of $^{50}$V. The energy difference between the hypernuclei
  and $^{50}$V  at $\beta=0$ is indicated with the numbers. Taken from Ref.~\cite{Xue15}.}
\end{figure}

Figure~\ref{fig:impurity} displays the energy surfaces for $^{51}_{~\Lambda}$V and its core nucleus $^{50}$V as a function of the quadrupole deformation parameter $\beta$ 
 from the deformed RMF calculation using the  PC-F1 \cite{Burvenich02}  parameterization for the $NN$ interaction and the
 PCY-S1 \cite{Tanimura12} parameterization  for the $N\Lambda$ interaction. 
The $\Lambda$ hyperon is always put in the lowest-energy states among those which are connected to the $s$, $p$, $d$ state in the spherical limit, respectively.   
It is seen that the energy minimum of hypernucleus $^{51}_{\Lambda s}$V is shifted
slightly towards spherical shape, while those of $^{51}_{\Lambda p}$V and $^{51}_{\Lambda d}$V are pushed to a larger deformed shape.
Moreover, it is shown that the deformation of hypernuclei increases from $^{51}_{\Lambda s}$V to $^{51}_{\Lambda p}$V, and then to $^{51}_{\Lambda d}$V.
The difference in the $\Lambda$ binding energy $B_\Lambda$ values of $^{51}_\Lambda$V by the spherical and deformed RMF calculations is also shown
clearly in Figure~\ref{fig:impurity}, where the $B_\Lambda$ decreases by 0.1 MeV  or increases by 0.8 MeV and 2.2 MeV for the $\Lambda_s$, $\Lambda_p$ and $\Lambda_d$, respectively, after considering deformation effect.

% Figure
\begin{figure}[h!]
\label{MFcurve}
 \includegraphics[width=200pt,height=200pt]{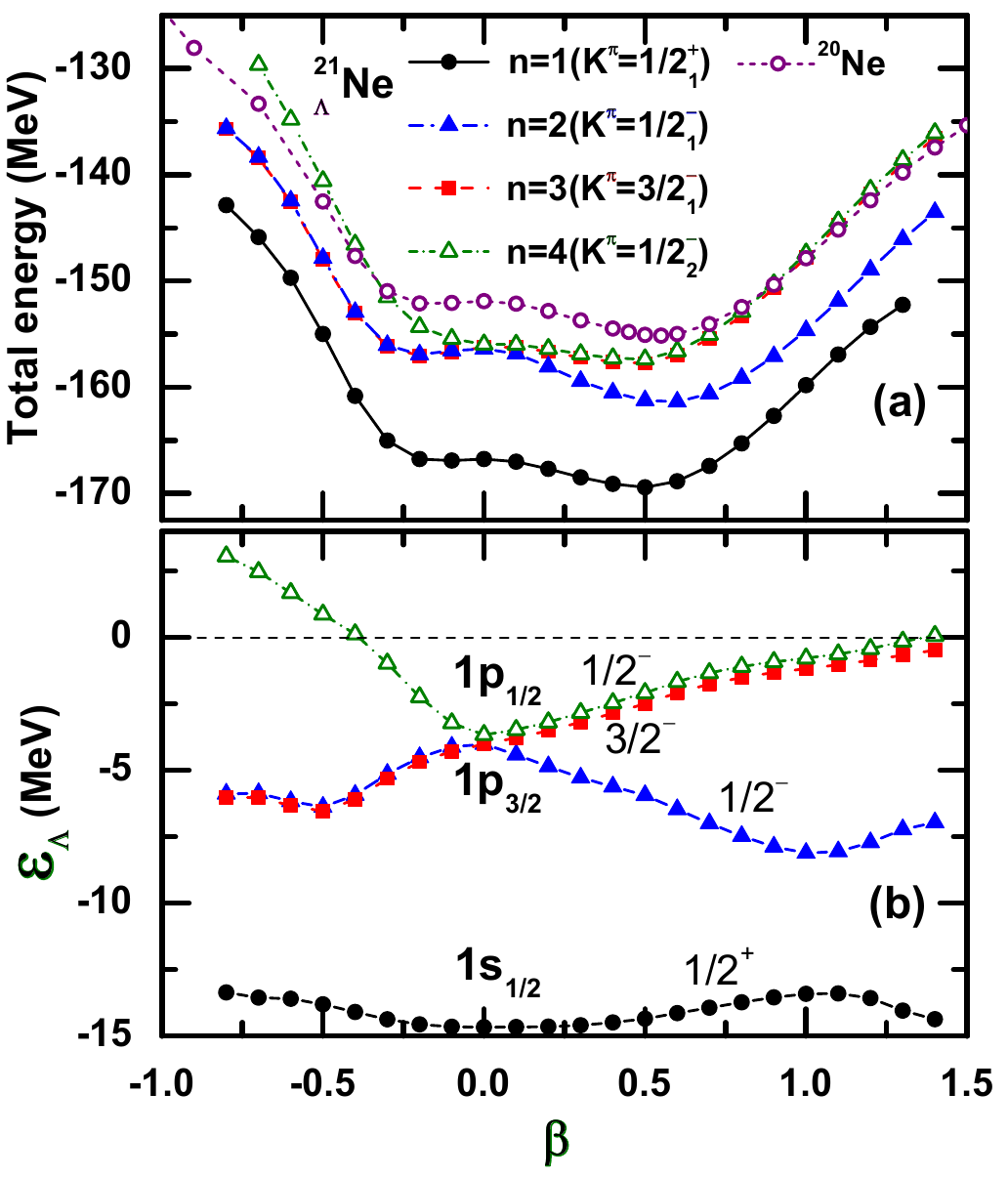}
 \caption{(Color online)(a) The total energy curves for $^{21}_\Lambda$Ne obtained in the mean-field approximation
as a function of quadrupole deformation $\beta$. These are calculated by putting
the $\Lambda$ hyperon in different single-particle orbitals
shown in the lower panel. For comparison,
the energy curve for the core nucleus $^{20}$Ne is also
plotted.  (b) The single-particle energies of the
$\Lambda$ hyperon in  $^{21}_\Lambda$Ne as a
function of quadrupole deformation.
These are labeled with the $\Omega^\pi$ number, that is
the projection of the angular momentum
onto the $z$-axis in the body fixed frame. Taken from Ref.~\cite{Mei16R}.}
\end{figure}

 \begin{figure}[]
  \centering
 \includegraphics[width=350pt]{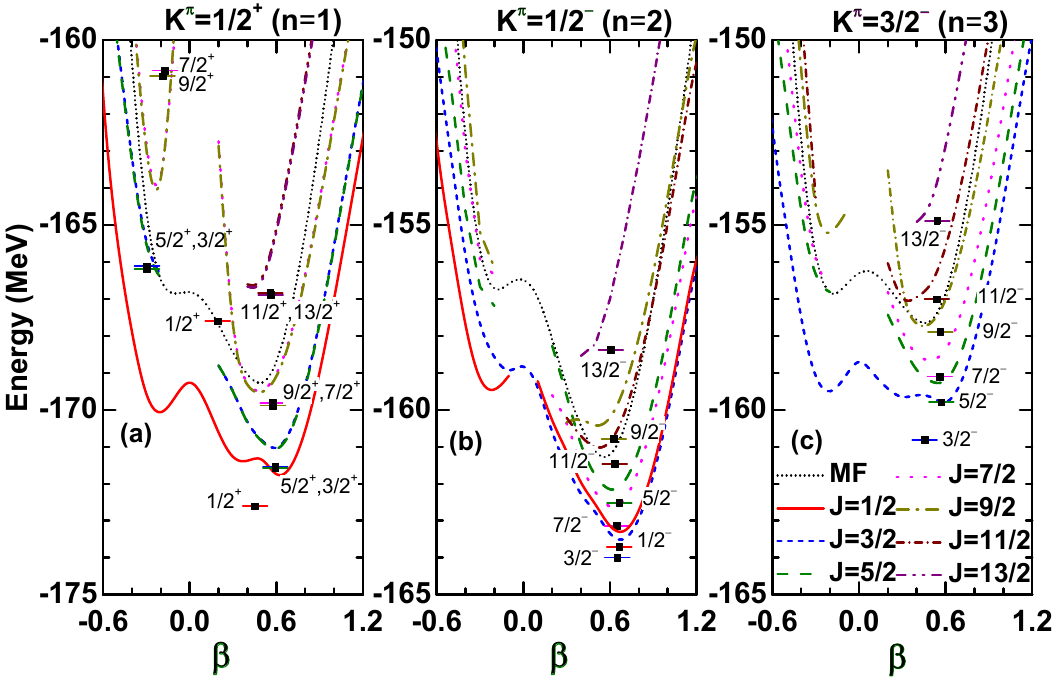}
 \caption{(Color online) The projected energy curves for 
$^{21}_\Lambda$Ne obtained by putting  
the $\Lambda$ hyperon on the three lowest single-particle orbitals 
labeled by $K^\pi$. 
The corresponding mean-field energy curves are also shown for a comparison. 
The solutions of the GCM calculations 
are indicated by the squares and the horizontal bars 
placed at the average deformation. Taken from Ref.~\cite{Mei16R}}
\label{ProjCurve}
\end{figure}

 \begin{figure}[]
  \includegraphics[width=250pt]{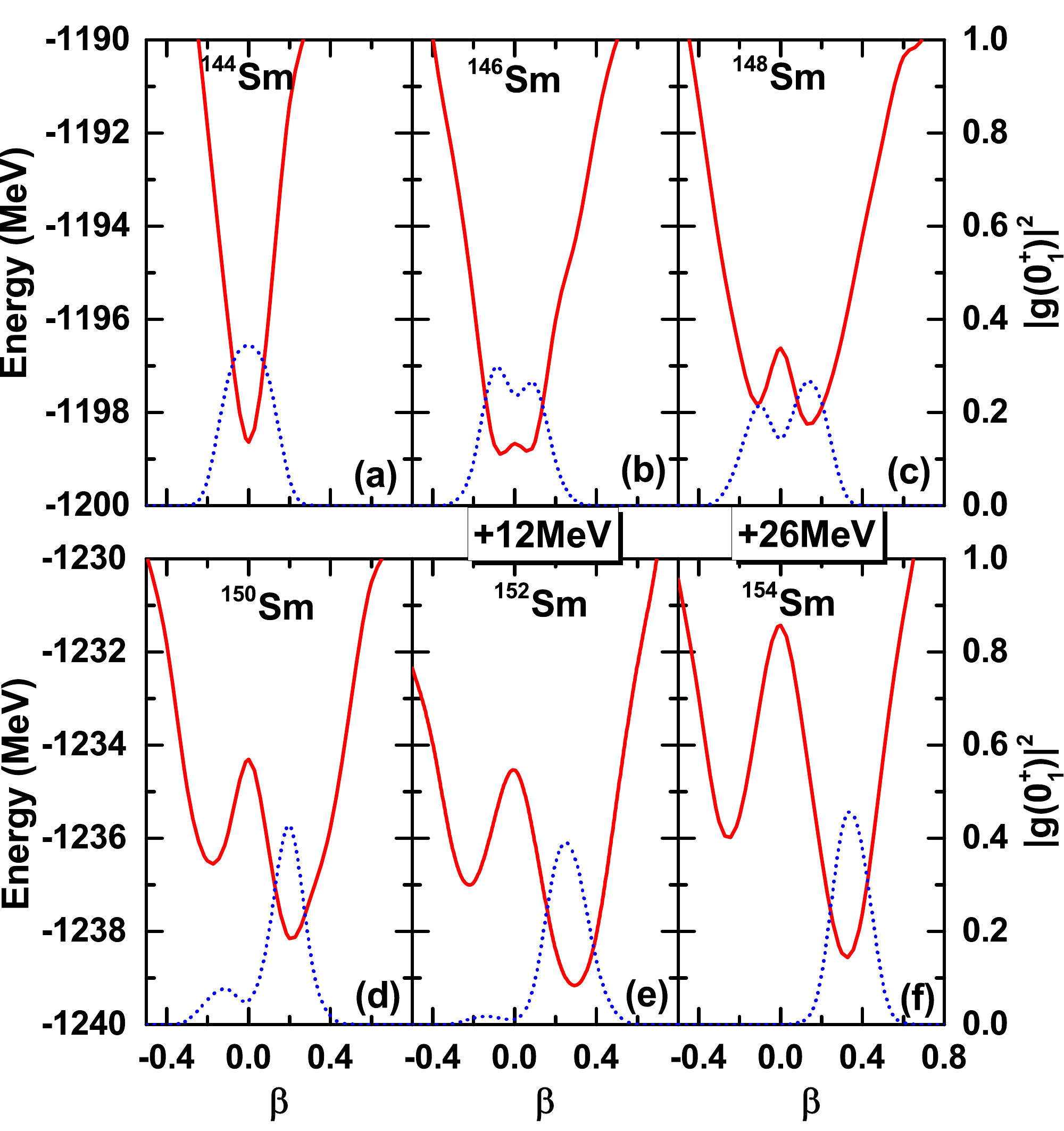}
  \caption{The total energy in the mean-field approximation for Sm
isotopes as a function of the the intrinsic
quadrupole deformation $\beta$ (the solid lines).
The energy curves for $^{146}$Sm and $^{152}$Sm
are shifted upward by 12 MeV, while those for
$^{148}$Sm and $^{154}$Sm are shifted by 26 MeV.
The square of the collective wave functions, $|g_{nI}(\beta)|^2$ with
$g^I_{n}(\beta)\equiv\sum_{\beta'} \big[ \mathcal{N}^{I}\big]^{1/2}(\beta,\beta')f^I_{n}(\beta')$, for
the GCM ground states ($0_1^+$) are also shown with the dashed curves. Taken from Ref.~\cite{Mei17}.}
  \label{Sm-PES}
\end{figure}

Figure~\ref{MFcurve}(a) shows the  energy surfaces for $^{21}_{~\Lambda}$Ne (with the $\Lambda$ in different orbit) and its core nucleus $^{20}$Ne.
In axially deformed case, one can use $K^\pi$  to label configurations. One can see that the energies for the three negative-parity configurations (that is, $K^\pi=1/2^-_1, 3/2^-_1$, and $1/2^-_2$), corresponding to the hyperon occupying the three ``$p$-orbital" states, are close to each other at $\beta=0$ due to a weak hyperon spin-orbit
interaction, and are well separated from the energy of the positive parity configuration ($K^\pi=1/2^+_1$), which
corresponds to the hyperon occupying the ``$s$-orbital" state. It is shown that the energy minimum appears at
$\beta\sim0.6$ for $K^\pi=1/2^-_1$, which is larger than the deformation of the energy minimum
for the $1/2^+_1$ configuration ($\beta=0.49$). It is consistent with the findings in $^{51}_\Lambda$V shown in Fig.~\ref{fig:impurity} . 
Figure~\ref{MFcurve}(b)  displays the  Nilsson diagram for the hyperon, from which, one sees clearly that the
energy of the lowest-energy $p$ orbital (labeled as $1/2^-$) is decreasing with the deformation (up to $\beta_2=1.0$). In other words,
 the $\Lambda$ on this orbit is generally energetically favored in deformed shape and thus has the deformation-driving effect.

Figure~\ref{ProjCurve} displays the projected energy curves of $^{21}_\Lambda$Ne as a function of $\beta$ obtained by taking the diagonal element of the Hamiltonian and the norm kernels as $E^J_n(\beta)={\cal H}^J_{n}(\beta,\beta)/{\cal N}^J_{n}(\beta,\beta)$.  Besides, the predicted low-lying states of $^{21}_\Lambda$Ne after mixing all the projected mean-field states for each $K^\pi$ configuration with the GCM method are indicated by the squares in the figures.  It is seen that the prolate minimum in the projected energy curves becomes more pronounced
and thus the nuclear shape becomes more stable as the angular momentum increases.
Moreover,  the energy minimum for the $J^\pi=1/2^+$ energy curve appears at
deformation $\beta=0.62$, that is somewhat larger than the deformation at the minimum of
the corresponding mean-field curve, $\beta=0.49$, due to the energy gain originated from the angular momentum projection.
On the other hand, if one compares it to the projected energy curve for the 0$^+$
configuration of $^{20}$Ne, which has a minimum at $\beta=0.65$,
one finds again that the minimum is slightly shifted towards the spherical
configuration both on the oblate and the prolate sides.

In contrast to the $J^\pi=1/2^+$ configuration, the deformation at the energy minimum for the
$J^\pi=1/2^-$ configuration increases to $\beta=0.69$ (see
Fig.~\ref{ProjCurve}(b)). Moreover, for this configuration, the energy difference
between the prolate and the oblate minima significantly increases
as compared to the $J^\pi=1/2^+$ configuration.
For this reason, the collective wave function for the $J^\pi=1/2^-$
state is expected to be more localized on the prolate side
than that of the $J^\pi=1/2^+$ state.
As a consequence, the average deformation for the $J^\pi=1/2^-$
state is close to the minimum point of the energy curve
while that for the $J^\pi=1/2^+$ configuration is shifted towards
the oblate side due to a cancellation between the prolate and the
oblate contributions (see the filled squares
in Fig.~\ref{ProjCurve}(a)
and \ref{ProjCurve}(b)). The projected energy curves for the $K^\pi=3/2^-_1$
configuration are shown in Fig.~\ref{ProjCurve}(c).
These are several MeV higher than those for the $K^\pi=1/2^-_1$
configuration. Besides, the energy curve for the $J^\pi=3/2^-$ is considerably different
from that for the $J^\pi=5/2^-$ configuration, and one would not
expect a (quasi-)degeneracy between these two states. 

We note that $^{20}$Ne has low-lying negative-parity states originated from the $\alpha+^{16}$O
cluster structure \cite{Zhou16}, which would also exist in $^{21}_\Lambda$Ne. 
It would be interesting to study how the octupole correlation affect the low-lying states of $^{21}_\Lambda$Ne
in the future.

%Currently, there is no corresponding spectroscopic data available for comparison.

\section{Collective correlations and configuration mixings in low-lying states of $\Lambda$ hypernuclei}

Figure \ref{Sm-PES} show the evolution of potential energy surfaces and collective wave function of the ground state in Sm isotopes around neutron number $N=90$.  It is exhibited clearly a picture of   shape transition from vibrational to rotational characters as the number of neutrons increases. Therefore, the  Sm isotopes provide an ideal playground to study how nuclear collective correlations change the configuration mixings in hypernuclear low-lying states.  

% Figure
\begin{figure}[]
  \includegraphics[width=200pt,height=200pt]{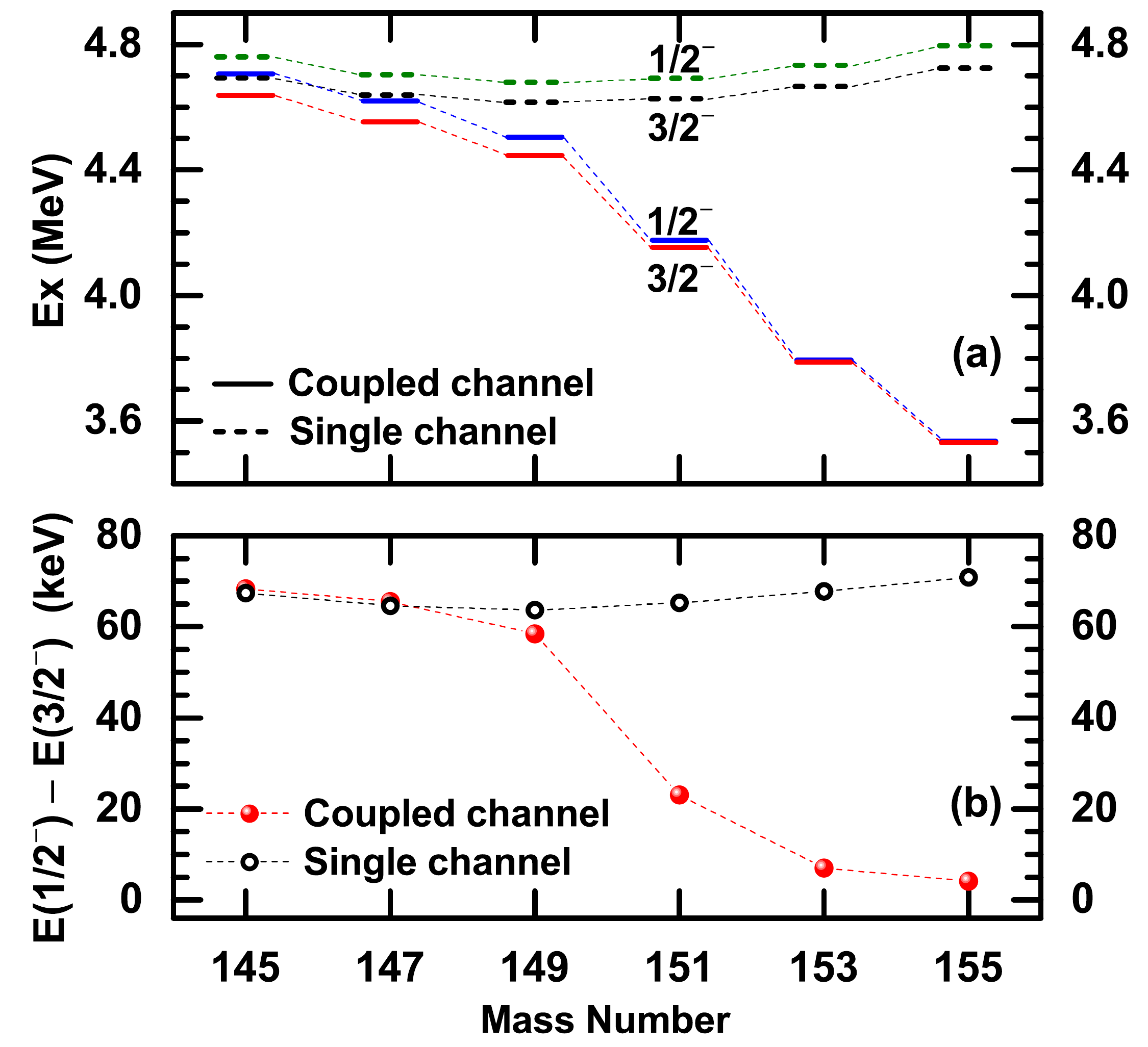}
   \includegraphics[width=200pt,height=200pt]{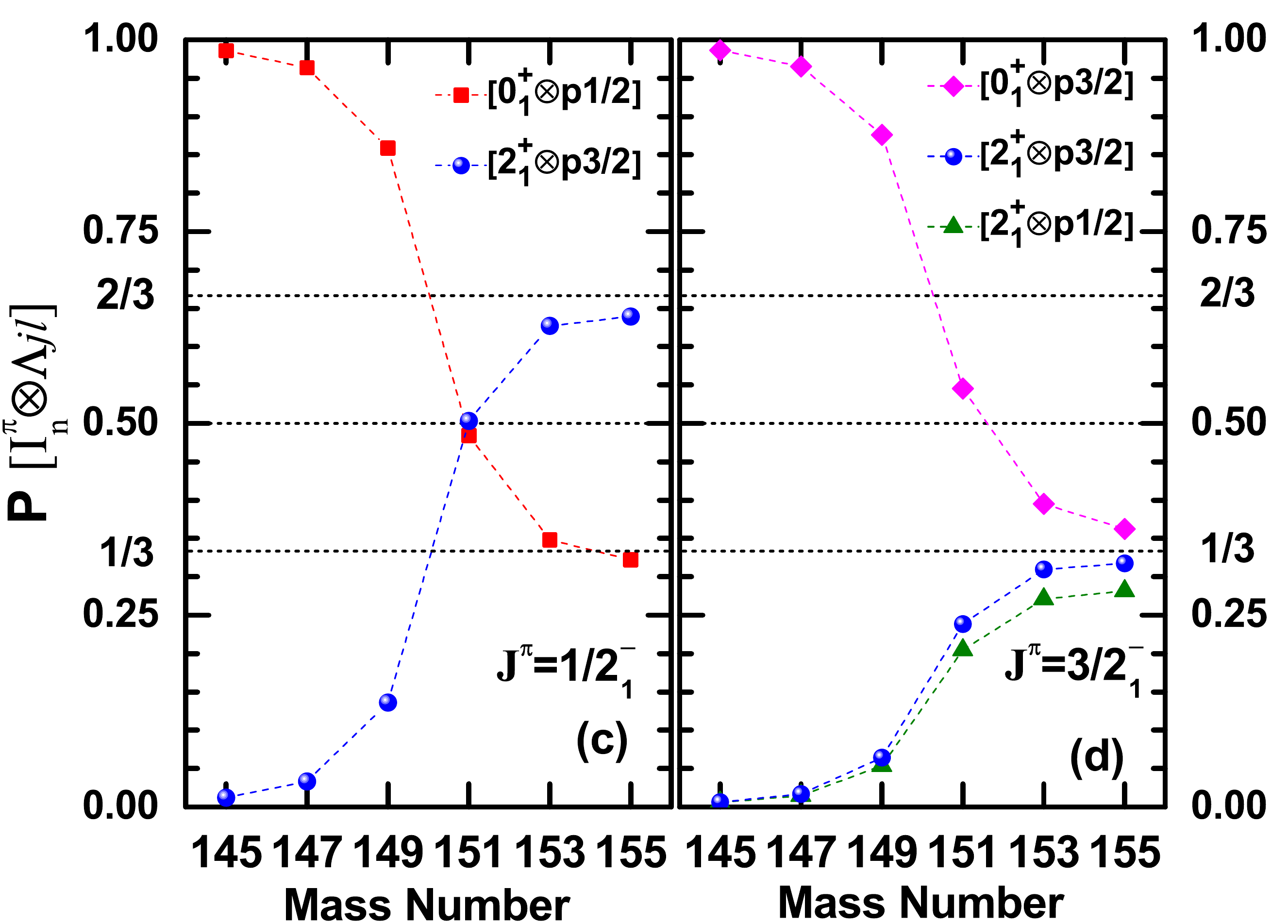} 
  \caption{(a) The energy levels of the $1/2^-_1$ and $3/2^-_1$ states in the Sm hypernuclei as a function of the mass number.   (b) The energy splitting between the $1/2^-_1$ and $3/2^-_1$ states.  The probability $P_k\equiv \int dr r^2 \vert {\cal R}_{j\ell n I}(r)\vert^2$ for the dominant components in the wave function of (c) the $1/2^-_1$ state and (d) the $3/2^-_1$ state as a function of the mass number of the $_\Lambda$Sm isotopes. Taken from Ref.~\cite{Mei17}.}
    \label{SplitDevelop}
\end{figure}

Figures~\ref{SplitDevelop}(a) and (b) show the excitation energy  of the lowest $1/2^-_1$ and $3/2^-_1$ states  in the Sm hypernuclei
as a function of the neutron number from the PCC calculation.  The dashed lines show the results of single-channel calculations, for  which the sum in Eq. (\ref{wavefunction}) is restricted only to a single  configuration.  For the lowest $1/2^-_1$ and $3/2^-_1$ states, 
the configuration in the single-channel calculation is  a pure configuration of $[ 0^+_1 \otimes \Lambda p_{1/2} ]$ 
and $[0^+_1 \otimes \Lambda p_{3/2} ]$, respectively.  Their excitation energies are around 4.8 MeV for all the hypernuclei 
considered in this paper, which is close to  the energy $\frac{2}{3}\times41A^{-1/3}\sim 5.14$ MeV with $A\sim 150$ 
for exciting one hyperon  from $s$ orbit to $p$ orbit. 
The energy difference between these states remains around 70 keV, as shown 
by the open circles in  Figure~\ref{SplitDevelop}(b). 
In contrast,  the energy of the $1/2^-_1$ and $3/2^-_1$ states 
obtained by including the configuration mixing effect 
decreases  continuously from 4.7 MeV to 3.5 MeV as the neutron number 
increases from 82 to 92 (see the solid lines in 
Fig.~\ref{SplitDevelop}). 
The splitting of these two states also decreases from 68 keV to 4 keV,
as shown in the left panel of Fig.~\ref{SplitDevelop} by the filled circles. 
The deviation from the single-channel calculations increases as 
the core nucleus undergoes phase transition 
from a spherical vibrator to a well-deformed rotor, indicating 
a stronger configuration mixing effect in deformed hypernuclei. 
 
The mass number dependence of the mixing amplitude is shown 
in Figs.~\ref{SplitDevelop}(c) and (d), indicating a similar feature as in the $1/2_1^-$ state. 
One can see that the configuration mixing becomes stronger as the core nucleus 
undergoes a transition from spherical to deformed shape. 
For  $^{155}_{~~\Lambda}$Sm, the weight factors are  36.3\%,  28.1\%, and 31.8\%, 
for the  $[0_1^+ \otimes \Lambda p_{3/2}]$, $[2_1^+\otimes \Lambda p_{1/2} ]$ 
and $[2_1^+\otimes \Lambda p_{3/2}]$ configurations, 
respectively.

\section{Spin-orbit splitting of $p$-orbital $\Lambda$ in carbon hypernuclei}

Based on the conclusion drawn from  the Sm hypernuclei, the energy splitting of the first $1/2^-$ and $3/2^-$  states in hypernuclei only with weak collective correlations 
can be safely interpreted as the spin-orbit splitting of $p_\Lambda$.   $^{13}_{~\Lambda}$C is a good candidate hypernucleus for this purpose as $^{12}$C is weakly deformed.
It has been proved by the microscopic cluster model calculation \cite{Hiyama00} for $^{13}_{\Lambda}$C, which shows that the  91.1\% (95.7\% ) of the $3/2^- (1/2^-)$ state is the configuration of  $[^{12}$C$(0^+)\otimes \Lambda p_{3/2}]$ ($[^{12}$C$(0^+)\otimes \Lambda p_{1/2}])$. 
The $\gamma$-rays from the excited $1/2^-_1$ and $3/2^-_1$ states to the ground state were measured following the $^{13}$C($K^-$,$\pi^-$)$^{13}_{~\Lambda}$C reaction. The energy difference between the $1/2^-_1$ and $3/2^-_1$ states was determined to be $152\pm54({\rm stat}) \pm36$({\rm syst})~keV~\cite{Ajimura01}, which was interpreted as the spin-orbit splitting between 1$p_{1/2}$ and 1$p_{3/2}$ hyperon states in $^{13}_{~\Lambda}$C.  
The neutron number in $^{14}$C is a magic number and thus the collective correlation in $^{14}$C is expected to be weaker than that in $^{12}$C. It is interesting to study configuration mixing in $^{15}_{\Lambda}$C.

\begin{table}
\caption{The excitation energies $E_x$ (in unit of MeV) of the lowest $1/2^-$ and $3/2^-$ states  and their energy splittings in carbon hypernuclei. Taken from Ref.\cite{Xia17}.}
\label{tab:splitting} 
\tabcolsep7pt
\begin{tabular}{lccccc}
\hline
 & $^{13}_{~\Lambda}$C &  $^{15}_{~\Lambda}$C &$^{17}_{~\Lambda}$C & $^{19}_{~\Lambda}$C \\
 \hline
   $E_x(1/2^-)$   &   12.964       &  12.224  &   10.498    & 10.027 \\
   $E_x(3/2^-)$   &   12.711       &  11.880   &   10.431   &  9.994 \\
\hline
${\rm splitting}$       &    0.253   & 0.344 &	0.067 &	0.033\\
 \hline
\end{tabular} 
\end{table}

% Table
\begin{table}
 \caption{The probability $P_k$ of the dominant components in the wave functions for some selected negative-parity states. The components with   probabilities
smaller than 0.001 are not given.  Taken from Ref.~\cite{Xia17}.}
 \label{table:Component}
\tabcolsep7pt
\begin{tabular}{lccccc}
  \hline\hline
$J^\pi$    &$(lj)\otimes I^\pi_{n}$ & $^{13}_{~\Lambda}$C &$^{15}_{~\Lambda}$C &$^{17}_{~\Lambda}$C &$^{19}_{~\Lambda}$C
\\     \hline
$1/2^-_1 $  &$p_{1/2}\otimes0_1^+$ &$0.8853$ &$0.9643$ &$0.5477$ &$0.5705$  \\
  $$        &$p_{3/2}\otimes2_1^+$ &$0.1054$ &$0.0339$ &$ 0.4400$ &$0.4066$  \\

$3/2^-_1 $  &$p_{1/2}\otimes2_1^+$ &$0.0453$ &$0.0148$ &$0.1913$ &$0.1865$      \\
            &$p_{3/2}\otimes0_1^+$ &$0.9207$ &$0.9738$ &$0.6071$ &$0.6184$  \\
            &$p_{3/2}\otimes2_1^+$ &$0.0243$ &$0.0089$ &$0.1897$ &$0.1682$     \\
 \hline\hline
 \end{tabular} 
\end{table}

Table~\ref{tab:splitting} and Table~\ref{table:Component} list the energies and dominant components of the  $1/2^-_1$ and $3/2^-_1$ states  in carbon hypernuclei, respectively. For $^{13}_\Lambda$C, the predicted energy difference between the two negative-parity states  $\Delta E$  is 0.253  MeV, close to the data $\Delta E=0.152(90)$ MeV~\cite{Hashimoto06}. For $^{15}_\Lambda$C, the energy difference between the $1/2^-_1$ and $3/2^-_1$ states is predicted to be 0.344 MeV, about 0.1 MeV larger than that in $^{13}_\Lambda$C. In contrast to the cases in $^{13,15}_{~~~~~\Lambda}$C, this value is only 67 keV and 33 keV in $^{17,19}_{~~~~~\Lambda}$C, respectively. One can see from Table~\ref{table:Component} that the energy splitting of the $1/2^-_1$ and $3/2^-_1$ states in $^{17,19}_{~~~~~\Lambda}$C cannot be interpreted as the spin-orbit splitting of the $p_\Lambda$ state due to the large configuration mixing.   In short, the results indicate that $^{15}_\Lambda$C is a more ideal hypernucleus than $^{13}_\Lambda$C to extract the $\ell s$ splitting of the $p_\Lambda$ state, even though the production of $^{15}_\Lambda$C on experiment is much more difficult.

\section{Uncertainty in the $N\Lambda$ interactions and its impact on neutron stars}
The coupling strengths of the effective $N\Lambda$ are often determined by fitting to the $\Lambda$ binding energy $B_\Lambda$ which is approximately 
given by the $\Lambda$ single-particle energy $B_\Lambda \approx -\epsilon_\Lambda$, where the single-particle energy of the $\Lambda$  in the RMF approaches
\beq
\epsilon_\Lambda =  \frac{p^2_\Lambda}{2m_\Lambda} + V_\Lambda+ S_\Lambda + {\cal O}(1/m^2_\Lambda)
\eeq
is mainly governed by the cancellation of attractive scalar  $S_\Lambda=x_\sigma g_{\sigma N}\sigma$ and repulsive vector $V_\Lambda=x_\omega g_{\omega N}\omega^0$ potentials~\cite{Xue15}.  As pointed out by Glendenning~\cite{Glendenning91}  that there is a continuous ambiguity in the pair  of values $(x_\sigma, x_\omega)$ which are able to reproduce the $\Lambda$ binding energy in nuclear matter. A similar phenomenon was also seen in finite hypernuclei \cite{Keil00,Wang13}. The ambiguity in the coupling strengths may cause a large uncertainty in the predicted maximum mass of neutron starts. It was shown in Ref. \cite{Long12} that with the weakening of $N\Lambda$ coupling gradually, more and more neutrons are transferred into $\Lambda$ hyperons, and the EOS will become increasingly softer.  Generally speaking, the weaker the $ N\Lambda$ coupling the lower the maximum mass of neutron stars. It was found that the predicted maximum mass of neutron stars can still reach the value $(1.97\pm0.04)M_\odot$ with the presence of hyperons by choosing sufficient large values ($x_{\sigma/\omega} > \sim 0.7-0.8$) for the coupling strengths \cite{Glendenning91,Long12,Sun18}, even though $x_{\sigma/\omega}=2/3$ is suggested by the naive quark model.  Therefore, a precise calibration of the  $N \Lambda$ interaction is highly important for understanding the so-called ``hyperon puzzle" in neutron stars within the RMF framework.  In this subsection, we discuss  whether the energies of hypernuclear low-lying excited states can provide additional constraints on the $N\Lambda$ interaction or not. In the analysis, the relativistic point-coupling  $N\Lambda$ interactions PCY-S1 ($x_\sigma\approx 0.53, x_\omega\approx 0.64$), PCY-S2 ($x_\sigma\approx 0.11, x_\omega\approx 0.05$), PCY-S3 ($x_\sigma\approx 0.53, x_\omega\approx 0.63$), PCY-S4 ($x_\sigma\approx 0.48, x_\omega\approx 0.58$) \cite{Tanimura12} are adopted.

\begin{figure}[h]
  \centering
 \includegraphics[width=7cm]{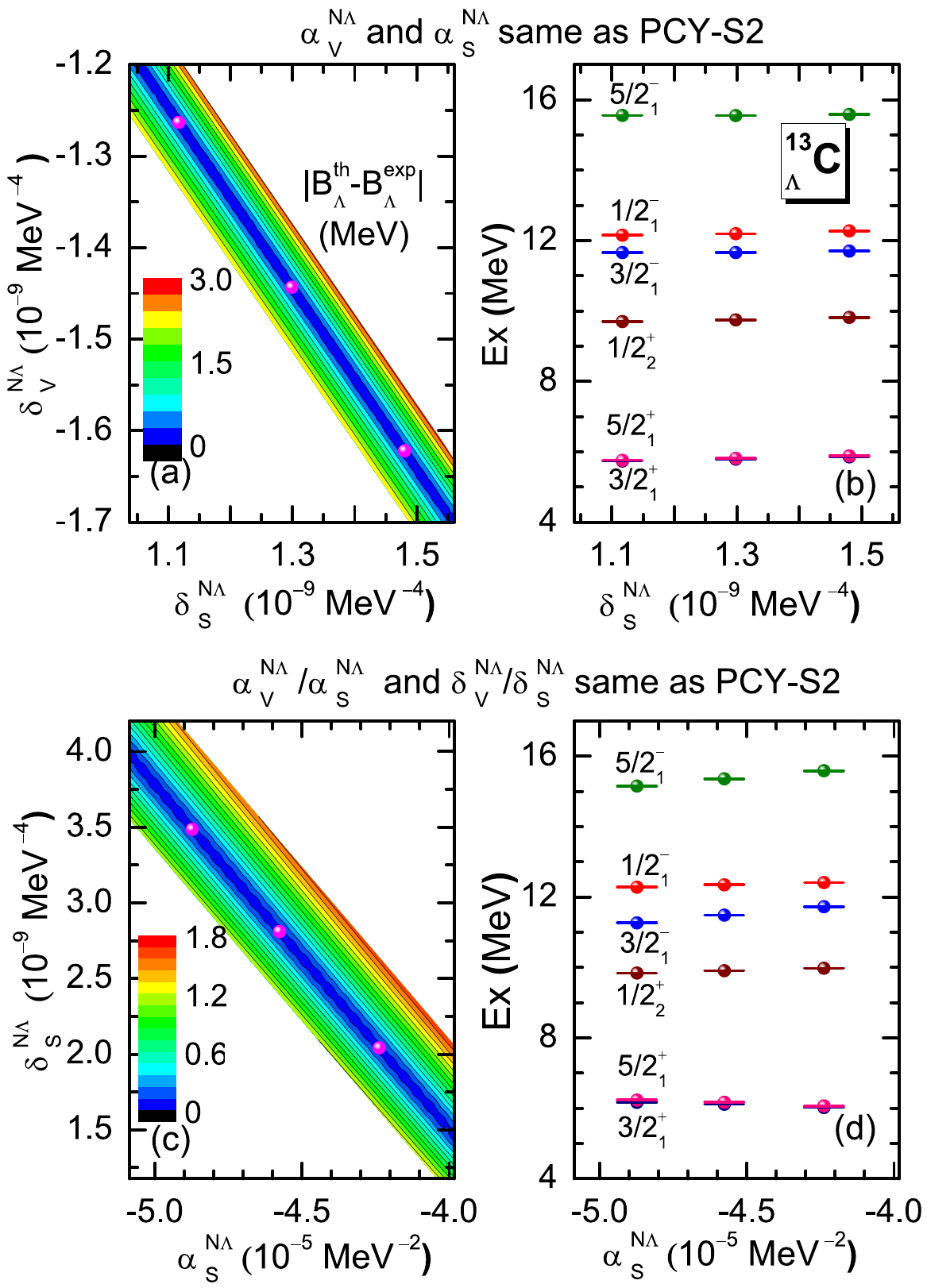} 
 \caption{(Color online) (a) and (c): Contour plots for the absolute value of the difference
between the
theoretical and the experimental hyperon binding energies of $^{13}_{~\Lambda}$C hypernucleus
as a function of the coupling strength parameters ($\delta^{N\Lambda}_S$, $\delta^{N\Lambda}_V$)
and ($\delta^{N\Lambda}_S$, $\alpha^{N\Lambda}_S$), respectively.
In the former, $\alpha^{N\Lambda}_V$ and $\alpha^{N\Lambda}_S$ are fixed to the same values
as in PCY-S2,
while in the latter, the value of $\alpha^{N\Lambda}_V$ and $\delta^{N\Lambda}_V$
is determined for each
($\alpha^{N\Lambda}_S,\delta^{N\Lambda}_S$) so as to keep the ratios
$\alpha^{N\Lambda}_V/\alpha^{N\Lambda}_S$ and $\delta^{N\Lambda}_V/\delta^{N\Lambda}_S$
to be the same as those for PCY-S2.
(b) and (d): Low-lying states
in $^{13}_{~\Lambda}$C calculated with the  strength parameters
denoted by the dots in the panels (a) and (c), respectively. Taken from Ref. \cite{Mei16}. }
   \label{S2LODer}
\end{figure}
 
By fixing the coupling strengths $\alpha^{N\Lambda}_V, \alpha^{N\Lambda}_S$  to be the same values as
those in the PCY-S2 parameter set \cite{Tanimura12},  we  study the $B_\Lambda$ as a function of the coupling strengths $\delta^{N\Lambda}_V$ and $\delta^{N\Lambda}_S$.  
The results are shown in Fig.~\ref{S2LODer}(a). A clear linear correlation is observed between $\delta^{N\Lambda}_V$ and $\delta^{N\Lambda}_S$.
By selecting three sets of $(\delta^{N\Lambda}_V, \delta^{N\Lambda}_S)$ along the valley in Fig.~\ref{S2LODer}(a), we calculate the low-lying states of $^{13}_{~\Lambda}$C and show them  in Fig.~\ref{S2LODer}(b). One can see that the energies of the low-lying positive-parity states are very robust against the change of the parameters along the valley in Fig.~\ref{S2LODer}(a). Next, by fixing the values of $\alpha^{N\Lambda}_V/\alpha^{N\Lambda}_S$ and
$\delta^{N\Lambda}_V/\delta^{N\Lambda}_S$, we calculate the $B_\Lambda$ as well as the low-lying energy spectrum
as a function of $\alpha^{N\Lambda}_S$ and $\delta^{N\Lambda}_S$,
as shown in Figs.~\ref{S2LODer}(c) and (d), respectively. One can see that the parameters $\delta^{N\Lambda}_S$ and $\alpha^{N\Lambda}_S$ are also linearly correlated.

\begin{figure}[h]
  \centering 
 \includegraphics[width=8cm]{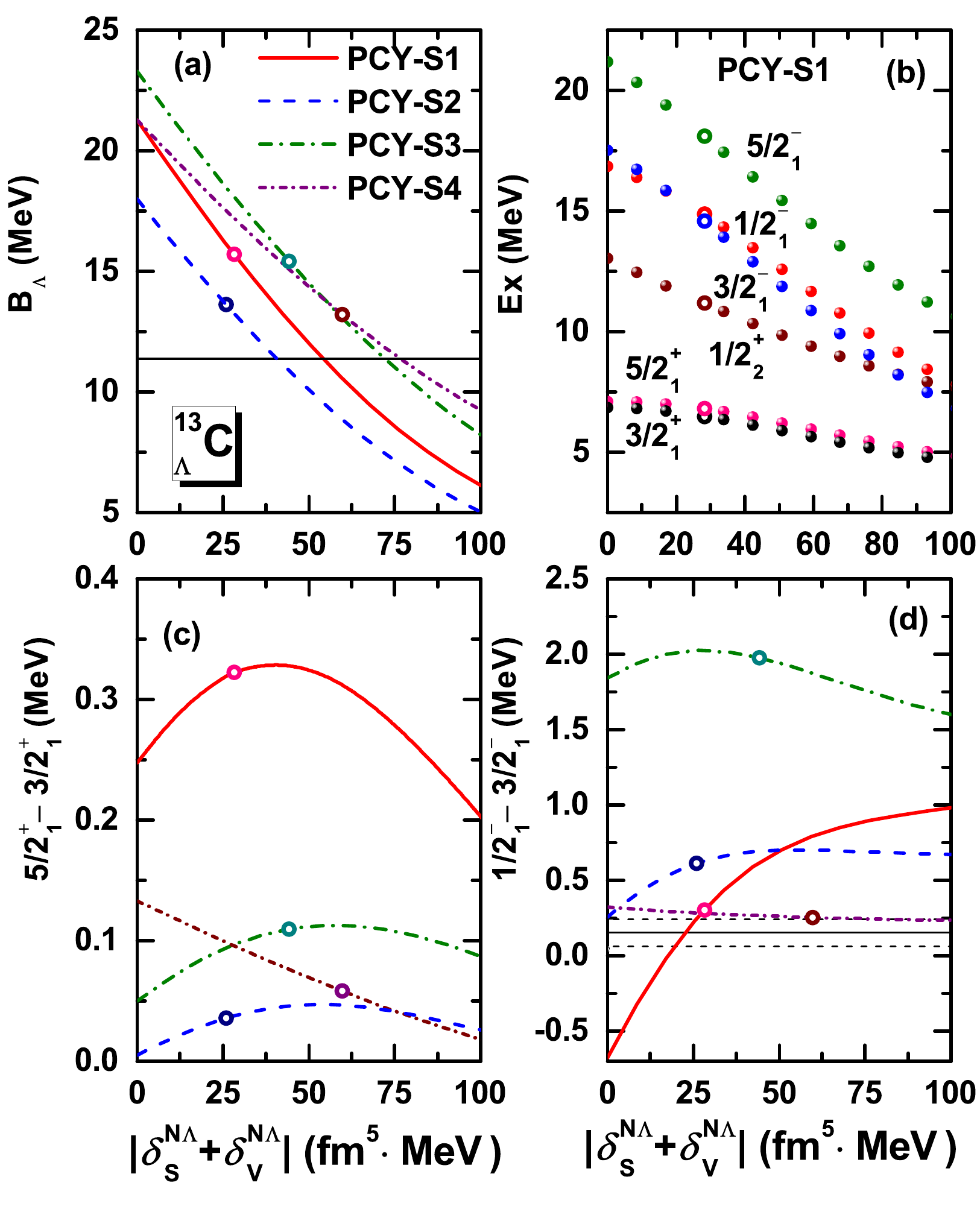}
  \includegraphics[width=8cm]{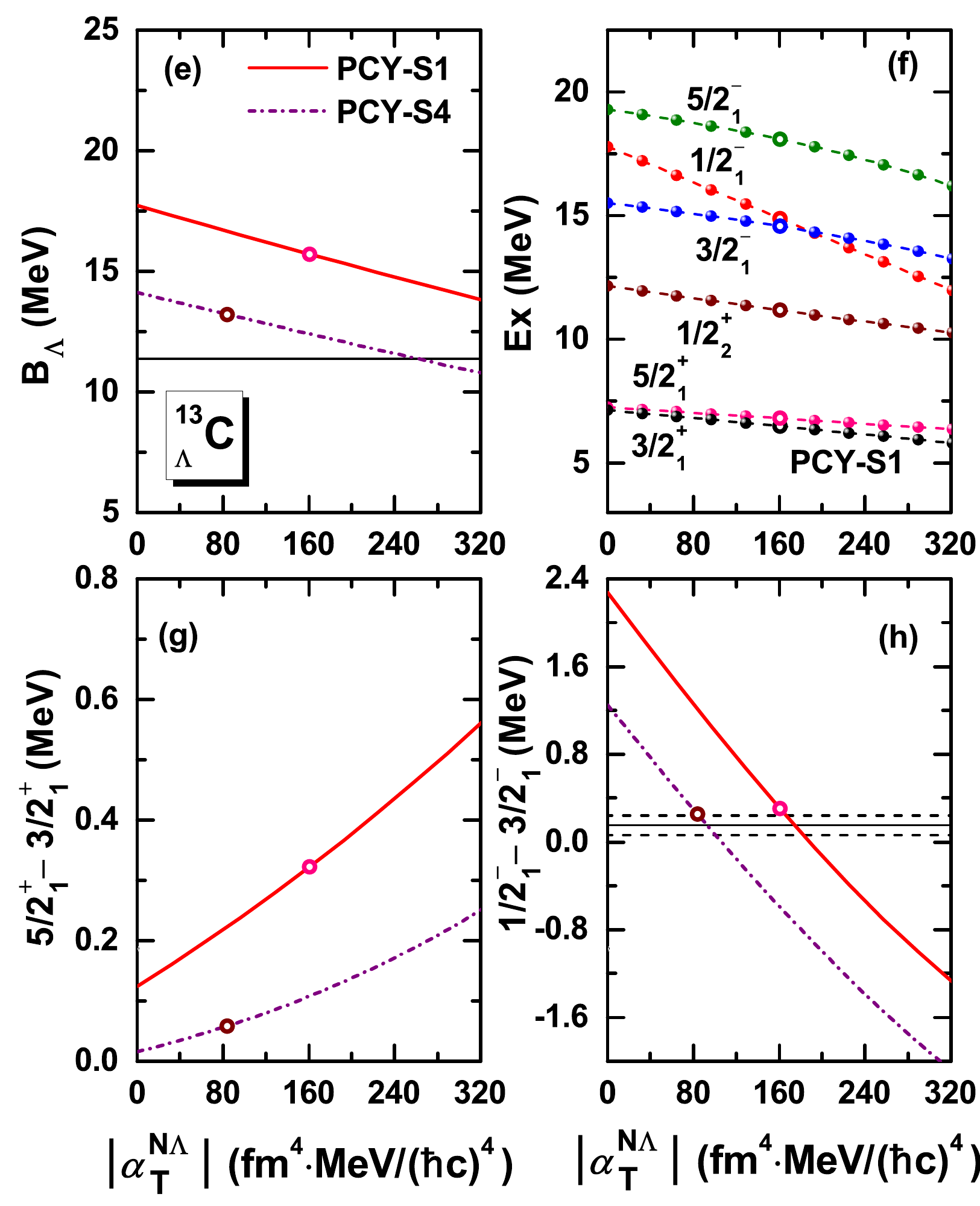} 
 \caption{(Color online) The $\Lambda$ binding energy (a), excitation energies of low-lying states (b), energy splitting of  
  $5/2^+_1$ and $3/2^+_1$ states (c) and that of  $1/2^-_1$ and $3/2^-_1$ states  (d)in $^{13}_{~\Lambda}$C as a function
of $|\delta^{N\Lambda}_S+\delta^{N\Lambda}_V|$.  (e)-(h) are for the same quantities, but as a function of   the tensor coupling strength
$|\alpha^{N\Lambda}_T|$.  The results by using the parameters of the PCY interactions are denoted by open circles . 
The experimental value is denoted by the thin solid line. Taken from Ref. \cite{Mei16}.
}
   \label{LODer}
\end{figure}

We next examine the influence of the derivative interaction terms for the other parameter
sets as well. To this end, we vary $\delta^{N\Lambda}_S$ and $\delta^{N\Lambda}_V$ by keeping the values of $\alpha^{N\Lambda}_S$, $\alpha^{N\Lambda}_V$, $\alpha^{N\Lambda}_T$ and the ratio $\delta^{N\Lambda}_S/ \delta^{N\Lambda}_V$
to be the same as the original values for each parameter set. The results are shown in Figs.~\ref{LODer}(a)-(d).
One can see that the $B_{\Lambda}$  decreases significantly with increasing $|\delta^{N\Lambda}_S+\delta^{N\Lambda}_V|$
and approaches to the experimental value denoted by the thin solid line.
The $\Lambda$ binding energy decreases from 21.28 MeV
to 15.72 MeV  by adding the derivative coupling terms to the PCY-S1 interaction
(that is, by changing $|\delta^{N\Lambda}_S+\delta^{N\Lambda}_V|$ from 0 to the original value
denoted by the open cicle).  The excitation energies also decreases with the increase of
$|\delta^{N\Lambda}_S+\delta^{N\Lambda}_V|$. Similar behaviors are found also for the PCY-S2, PCY-S3 and PCY-S4 forces (not shown).
The $5/2^+_1$ state is always slightly higher than  the $3/2^+_1$ state,
which is  by less than 0.15 MeV except for PCY-S1
in the range of $|\delta^{N\Lambda}_S+\delta^{N\Lambda}_V|$ shown in the figure. 
In short, the energies of low-lying states are sensitive to the the sum of the $\delta^{N\Lambda}_S$ and $\delta^{N\Lambda}_V$, which is also consistent with
the findings in Fig.~\ref{S2LODer}(a). In other words, there also exists a strong correlation between the strengths for the scalar and vector types of derivative couplings.

The impact of the tensor coupling term on hypernuclear low-lying states of $^{13}_{~\Lambda}$C 
is demonstrated in  Figs.~\ref{LODer}(e)-(h).
The $\Lambda$ binding energy gradually decreases from 17.71 MeV (14.12 MeV) for
$\alpha^{N\Lambda}_T=0$ to 15.72 MeV (13.22 MeV)
for the original value of $\alpha^{N\Lambda}_T$ for the  PCY-S1 (PCY-S4) force, which
is indicated by the open circle in Fig.~\ref{LODer}(e). The tensor coupling term makes the $\Lambda_{s1/2}$ hyperon less bound by increasing the energy of the $s_{1/2}$ level. 
Moreover, it decreases (increases) the energy of the hyperon $p_{3/2}$ ($p_{1/2}$)  state. As a result, the tensor coupling term decreases
(increases) the energy of the $3/2^-$ ($1/2^-$) state, which  mainly consists of
the $p_{3/2}$ ($p_{1/2}$) hyperon coupled to the ground state ($0^+$) of $^{12}$C.
Since the $1/2^-$ changes more significantly than the $3/2^-$ state, the higher
lying $1/2^-$  state approaches the $3/2^-$ state and even becomes lower than the $3/2^-$
state for large values of the tensor coupling strength,
indicating that the energy splitting of the $1/2^-$ and $3/2^-$ states is sensitive
to the tensor coupling strength.

\section{Summary and outlook}

 We have established relativistic mean-field (RMF) and beyond approaches for the low-lying states  of deformed hypernuclei.  The impurity effect of 
 $\Lambda$ hyperon at different orbits has been demonstrated. In particular, the collective correlations and configuration mixing in hypernuclear low-lying  states 
 have been examined in detail. Finally, we studied the sensitivity of the  energies of hypernuclear low-lying states to parameters in the effective $N \Lambda $ interaction.
 Strong correlations between the interaction parameters are exhibited.   Only after resolving the uncertainty in the interaction parameterizations can one have a solid understanding on the issue of so-called ``hyperon puzzle" in neutron stars. Further investigations on the effect of three-body $NN\Lambda$ coupling terms, the sensitivity of the electromagnetic transition strengths  to the coupling strengths,  and their impacts on  the predicted mass-radius relation of neutron stars are to be done in the future.  

% Acknowledgement
\section{ACKNOWLEDGMENTS}
 
We are indebted to our collaborators on the nuclear covariant density functional theory for hypernuclei at various stages, namely Z.P. Li, H.F. Lu, J. Meng, P. Ring, C.Y. Song, Y. Tanimura, X.Y. Wu, H.J. Xia, W.X. Xue,  and X.R. Zhou. We also thank T. Koike, H. Tamura  for many useful discussions. This publication is based on work supported in part by the Tohoku University Focused Research Project ``Understanding the origins of matter in the universe", JSPS KAKENHI Grant No. 2640263, and by the National Natural Science Foundation of China under Grant No. 11575148. JMY also acknowledges the support by the Scientific Discovery through Advanced Computing (SciDAC) program funded by the U.S. Department of Energy, Office of Science, Office of Advanced Scientific Computing Research and Office of Nuclear Physics, under Award Number DE-SC0008641 (NUCLEI SciDAC Collaboration).

% References

%\begin{thebibliography}{} 

\nocite{*}
\bibliographystyle{aipnum-cp}  
  \bibliography{./article}
 %\end{thebibliography}

%\bibliography{sample}%

\end{document}